\documentclass[12pt]{article}
\usepackage[latin1]{inputenc}
\usepackage{amssymb,graphicx}
\usepackage{epstopdf}
\usepackage{amsmath,amsfonts}
\usepackage{epsfig}

\def\dfrac#1#2{{\displaystyle#1\over\displaystyle#2}}

\begin{document}

\title{CMB anisotropy induced by \\ tachyonic perturbations of dark energy}

\author{M.V.~Libanov$^{\ddagger}$, V.A.~Rubakov$^{\ddagger}$,
O.S.~Sazhina$^{\star}$ and M.V.Sazhin$^{\star}$\\
\\
\\
$^{\dagger}$ Institute for Nuclear Research of the Russian Academy of
Sciences,\\
60th October Anniversary Prospect, 7a, Moscow, Russia;\\
$^{\star}$ P.K.~Sternberg Astronomical Institute of the Moscow State
University, \\
Universitetsky Prospect, 13, Moscow, Russia. }

\maketitle

\begin{abstract}

We study effects of possible tachyonic perturbations of dark energy on the
CMB temperature anisotropy. Motivated by some models of phantom energy, we
consider both Lorentz-invariant and Lorentz-violating dispersion relations
for tachyonic perturbations. We show that in the Lorentz-violating case,
the shape of the CMB anisotropy spectrum generated by the tachyonic
perturbations is very different from that due to adiabatic scalar
perturbations and, if sizeable, it would be straightforwardly
distinguished from the latter. The tachyonic contribution improves
slightly the agreement between the theory and data; however, this
improvement is not statistically significant, so our analysis results in
limits on the time scale of the tachyonic instability. In the
Lorentz-invariant case, tachyonic contribution is a rapidly decaying
function of the multipole number $l$, so that the entire observed dipole
 may be generated without conflicting the data at higher multipoles. On
the conservative side, our comparison with the data places limit on the
absolute value of the (imaginary) tachyon mass in the Lorentz-invariant
case.

\end{abstract}


\section{Introduction}
Recently, a number of suggestions have been put forward for explaining the
observed accelerated expansion of the Universe. Among them there are the
 presence of the cosmological constant, modification of gravity at
ultra-large scales, existence of new light fields (for reviews, see, e.g.,
Refs.~\cite{Padmanabhan:2002ji, Sahni:2004ai, Copeland:2006wr,
Sahni:2006pa, Frieman:2008sn}). In the latter case, dark energy can be
characterized by equation of state $p=w\rho$, where the parameter $w$ is
different from $-1$ and generically depends on time. In a simple version,
dark energy is due to a scalar field (quintessence), and the parameter $w$
obeys the constraint $w>-1$, while $w=-1$ for the cosmological constant.
However, the value of $w$  may be strongly negative, $w<-1$; this is the
case for phantom dark energy. Existing data do not exclude also a
possibility ~\cite{Komatsu:2008hk, Sahni:2008xx, Xia:2008ex} that at
relatively large redshift  the dark energy is of quintessence type $w>-1$,
and later it becomes phantom with $w<-1$.

Phantom energy violates null energy condition, which is usually a signal
for instabilities. As an example, the simplest model of  scalar field with
the wrong sign of kinetic term~\cite{Caldwell:2002} suffers from the
presence of ghost (negative energy state) at arbitrarily high spatial
momenta. This implies catastrophic vacuum instability. However, in phantom
models that break Lorentz-invariance, violation of the null energy
condition at cosmological scales (related to the property that $w<-1$ for
 spatially homogeneous phantom field) does not necessarily imply
unacceptable instabilities at shorter scales. This suggests that
Lorentz-violating phantom theories may be viable. Indeed, models of this
sort have been recently constructed~\cite{Senatore,
Crem:2006,Rubakov:2006, Libanov:2007}. A property of one class of these
models~\cite{Rubakov:2006, Libanov:2007} is that there is a tachyonic mode
in the perturbation spectrum about the homogeneous phantom background.
This mode  occurs at sufficiently small spatial momenta only, so that the
time scale of the tachyonic instability may be roughly comparable to the
 age of the Universe. This is not particularly dangerous.

It is conceivable that the existence of  tachyonic modes at low spatial
momenta is a fairly generic property of a class of phantom models: the
violation of the null energy condition may show up precisely in this way.
Therefore, it is of interest to study observable consequences of such
models. In this paper we consider one of these consequences, namely, the
effect of tachyonic modes on the anisotropy of CMB temperature. We adopt
the phenomenological approach, and instead of using results obtained
within a concrete model, we parametrize the tachyonic instability by a few
parameters. We will consider models  with a Lorentz-violating dispersion
relation and Lorentz-invariant one.

In the Lorentz-violating case, we parametrize the dispersion relation as
follows:
\begin{equation}
\omega^2 = \alpha |{\bf p}| ( M - |{\bf p}|),
\label{dispers}
\end{equation}
where $\alpha$  and $M$ are constant parameters, and ${\bf p}$ is the
physical spatial momentum. Our convention is that positive values of
$\omega ^{2}$ correspond to exponential growth of perturbations, while the
usual oscillatory behaviour occurs at negative $\omega ^{2}$. Thus, the
parameter $M$ equals to the momentum below which the mode is tachyonic.
For given $\alpha$, the parameter $M$ determines also the time scale of
instability. The parametrization (\ref{dispers}) is chosen in accord with
Refs.~\cite{Rubakov:2006, Libanov:2007} where similar dispersion relation
has been found in a concrete model of phantom energy. The analysis
presented below can be straightforwardly generalized to other forms of
dispersion relation.

In the Lorentz-invariant case the dispersion relation has the form
\begin{equation}
\omega ^{2}=M^{2}-\mathbf{p}^{2}
\label{Eq/Pg3/1:PhanTach}
\end{equation}
In this case too, the tachyon mass $M$ equals to the spatial momentum
below which the mode is unstable, and $M^{-1}$ is the time scale of
instability. From our perspective, the important difference between the
dispersion relations (\ref{dispers}) and (\ref{Eq/Pg3/1:PhanTach}) is that
in the latter case the ``frequency'' is nonzero at ${\bf p}=0$ and
monotonously decreases as $|{\bf p}|$ increases, whereas in the former,
the ``frequency'' vanishes as  ${\bf p}=0$ and has a maximum at finite
$|{\bf p}|$. This will lead to qualitatively different shapes of  the
contributions to the CMB anisotropy spectra. It is worth noting in this
regard that the Lorentz-invariant model may be viewed as a representative
of a class of theories with tachyonic perturbations: analogous results
would hold for Lorentz-violating models with dispersion relations similar
to (\ref{Eq/Pg3/1:PhanTach}).

In the cosmological context, the dispersion relations (\ref{dispers}) and
(\ref{Eq/Pg3/1:PhanTach}) are written as follows:
\begin{eqnarray}
\omega^2(t) &=& \alpha \dfrac{k}{a(t)}\left( M - \dfrac{k}{a(t)}\right) \;
, \nonumber \\
\omega^2 (t) &=&  M^2 - \frac{k^2}{a^2(t)} \; , \nonumber
\end{eqnarray}
respectively, where $k$ is the time-independent conformal momentum and
$a(t)$ is the scale factor. In the expanding Universe, a mode of given $k$
is first normal, and after the physical momentum gets reshifted down to
$k/a=M$, it becomes unstable.

For $ M  >  H_{0}$, where $H_{0}$ is the present value of the Hubble
parameter, the tachyonic modes in both models start to grow exponentially
at times preceding the present cosmological epoch. The growth of the
tachyonic perturbations gives rise to the growth of the gravitational
potential $\Phi$ generated by these
perturbations\footnote{Refs.~\cite{Libanov:2007, rub08} consider the
Lorentz-violating case, but that analysis is straightforwardly repeated in
the Lorentz-invariant situation.}~\cite{rub08}
\begin{eqnarray}
\Phi (t,\mathbf{x})&=&\frac{1}{(2\pi )^{3/2}}\int
\limits_{}^{}\!d\mathbf{k}\Phi (t,\mathbf{k})\mathrm{e}^{i\mathbf{kx}} +
\mbox{h.c.}, \nonumber\\
\Phi(t, {\bf k}) &=&{A({\bf k} )}
\exp\left(\int\limits^t_{t_k}\omega(t^{\prime})\;dt^{\prime}\right),
\label{gravpot}
\end{eqnarray}
where $A( {\bf k} )$ is the amplitude of primordial fluctuations in the
mode with conformal momentum ${\bf k}$ at the time $t_k$ when this mode
becomes unstable. We will discuss the range of  the amplitudes
$A(\mathbf{k})$ later on. We stress that the gravitational potential
(\ref{gravpot}) is generated by tachyonic perturbations rather than
inhomogeneities in the ordinary matter.

In this paper we calculate CMB multipoles generated by the gravitational
potential (\ref{gravpot}). In the Lorentz-violating case, the mechanism we
discuss gives rise to the contribution to the CMB anisotropy spectrum
which is quite different from the standard spectrum coming from adiabatic
scalar perturbations generated, e.g., at inflationary stage (for the
latter see, e.g., Refs.~\cite{book1, book2, book3}). As we will see below,
the tachyonic contribution leads to potentially observable features in the
CMB spectrum at relatively low multipoles. On the other hand, there are
hints towards the existence of deviations in the observed
spectrum~\cite{Komatsu:2008hk, Nolta:2008ih} from the predictions based on
   flat (Harrison-Zeldovich) or almost flat spectrum of primordial
adiabatic perturbations. So, one is tempted to employ tachyonic
instabilities to explain these deviations. We add the contribution due to
the tachyonic perturbations to the standard contribution of the adiabatic
modes and compare the result with the observed spectrum. We show that the
tachyonic contributions slightly improve the agreement between the theory
and data, but this improvement is statistically insignificant. So, our
analysis only leads to  limits on the parameters of the tachyonic
perturbations.

In the Lorentz-invariant case, tachyonic perturbations contribute to the
lowest multipoles only. An interesting possibility here is that they may
generate  the entire observed dipole without getting in conflict with
measurements of higher multipoles.

This paper is organized as follows. We begin with preliminaries on the
cosmological model in section~\ref{sec:background}, and then discuss in
detail the growth of the gravitational potential due to the tachyonic
modes in section~\ref{sec:growth}, first in the Lorentz-violating model
and then in the Lorentz-invariant one. In section~\ref{sec:multipoles} we
calculate the contributions to the CMB multipoles, again distinguishing
Lorentz-violating and Lorentz-invariant cases. We compare our results with
the data in section~\ref{sec:data}, and conclude in
section~\ref{sec:conclusion}.

\section{The cosmological model}
\label{sec:background}

The background space-time we consider in this paper corresponds to the
``almost'' standard cosmological model with the only special feature being
that the accelerated expansion of the Universe is driven by phantom energy
instead of the cosmological constant. The background metric is that of the
spatially flat expanding Universe,
\[
ds^2 = dt^2 - a^2(t)d\mathbf{x}^2.
\]
The scale factor $a(t)$ is determined by the Friedmann equation, which
can be written as follows:
\begin{equation}
\left( \frac{\dot a(t)}{a(t)} \right)^{2}=H_{0}^{2} \left[ \Omega_m \left(
\frac{a(t_0)}{a(t)} \right)^3 + \Omega_p \left( \frac{a(t)}{a(t_0)}
\right)^{-3(1+w_{p})}  \right],
\label{fried1}
\end{equation}
where $H_{0}$ is the present value of the Hubble parameter, $a(t_0) = a_0
= 1$ is the present value of the scale factor, the dot denotes the
derivative with respect to cosmic time $t$, $\Omega_p$ and $w_p$ refer to
phantom energy. We assume for simplicity that $w_p \equiv p_p/\rho_p$ is
independent of time; we will see in what follows that the effects we
discuss are largely independent of $w_p$, so this assumption is not
restrictive. The values we use in this paper are $\Omega_m=0.27$,
$\Omega_p=0.73$. According to observational data~\cite{Komatsu:2008hk} the
parameter $w_p$ belongs to the interval $-1.38 < w_{p} < -0.86$.

When calculating the CMB multipoles, we work with  conformal time $\eta$
instead of cosmic time $t$,
\[
\eta(t) = \int\limits_0^t \dfrac{d\hat t}{a(\hat t)}.
\]
Equation (\ref{fried1}), written in terms of conformal time, has the
form
\[
H_0 d\eta = \dfrac{ da}{\sqrt{a}\sqrt{\Omega_m + \Omega_p a^{-3w_p}}}.
\]

\section{Growth of perturbations}
\label{sec:growth}

The gravitational potential of a mode in the tachyonic regime is given by
eq.~(\ref{gravpot}). The exponent
\begin{equation}
N(t, k) = \int\limits_{t_k}^t \; d\hat t \; \omega(\hat t)
\label{Eq/Pg5/1:PhanTach}
\end{equation}
determines the growth of the potential in both Lorentz-violating model
(\ref{dispers}) and Lorentz-invariant one (\ref{Eq/Pg3/1:PhanTach}).
Properties of this function are different in the two cases, however.

\subsection{Lorentz-violating model.}
\label{sub:lv-growth}

In the Lorentz-violating model, the function (\ref{Eq/Pg5/1:PhanTach}) can
be written as an integral over the scale factor,
\begin{equation}
N(a, k) = \sqrt{\alpha }\frac{M}{H_{0}}\sqrt{\nu }\int\limits_{a_k}^a
d\hat a\dfrac{\sqrt{\hat a-\nu}}{\sqrt{\hat a}\sqrt{\Omega _{m} + \Omega
_{p} \hat a^{-3w_p}}},
\label{fold}
\end{equation}
where we have introduced dimensionless wave number
\[
{\bf \nu } = \dfrac{k}{M}.
\]
A few comments are in order. First, for a given mode the tachyonic
regime begins when the integrand in (\ref{fold}) becomes real, i.e., at $a
= \nu$. Since $a \leq a_0 =1$, the maximum value of $\nu$ for tachyonic
modes is $\nu =1$, that corresponds to the modes that are entering the
tachyonic regime today. In fact, because of the Hubble friction, the
exponential growth of the tachyonic mode starts not quite at the time when
$a=\nu$, namely, it starts at the time $t_k$ when the tachyonic
``frequency'' becomes comparable to the Hubble parameter,
\begin{equation}
\omega (t_{k}) \simeq H(t_{k}).
\label{Eq/Pg6/1A:Phantom}
\end{equation}
The lower limit of integration in (\ref{fold}) is the value of the scale
factor at that time, $a_k = a(t_k)$.

Second, the integral in (\ref{fold}) is a dimensionless smooth function of
its arguments $\{a,\nu \}$, both of which do not exceed 1. Therefore, this
integral is not parametrically large or small. On the other hand, the
amplitude $A(\mathbf{k})$ is small (see below), so the effect of the
tachyonic instability may be considerable only if $N(a,k)$ is sufficiently
large. This can happen if
\begin{equation}
\sqrt{\alpha }M>H_{0}.
\label{Eq/Pg6/1:Phantom}
\end{equation}
We assume in what follows that this inequality is indeed valid.

Third, the integral in (\ref{fold}) is saturated near its upper limit, and
because of (\ref{Eq/Pg6/1:Phantom}) this integral is practically
insensitive to the lower limit of integration. Therefore, it is an
excellent approximation to set $a_{k}=\nu $, that is, approximate the time
$t_k$ by the time at which the dispersion relation becomes tachyonic. In
fact, by solving eq.~(\ref{Eq/Pg6/1A:Phantom}) numerically, we have found
that $a_{k} = \nu $ with precision of order $10^{-5}$ in a wide range of
values of $\nu $ ($1\geq \nu \geq 0.05$). By setting $a_{k}=\nu $ in
(\ref{fold}) one finds that the dependence of $N$ on the parameters $M$
and $\alpha $ factors out
\begin{eqnarray}
N(a,\nu )&=&\sqrt{\alpha }\frac{M}{H_{0}}\mathcal{ N}(a,\nu )\; ,
\nonumber\\
\mathcal{ N}(a,\nu )&=&\sqrt{\nu }\int\limits_{\nu }^a d\hat
a\dfrac{\sqrt{\hat a-\nu}}{\sqrt{\hat a}\sqrt{\Omega _{m} + \Omega _{p}
\hat a^{-3w_p}}}\; . \nonumber
\end{eqnarray}

Fourth, the function $\mathcal{ N}(a, \nu)$ is obviously a growing
function of the scale factor $a$,
\begin{figure}[htb!]
\includegraphics[]{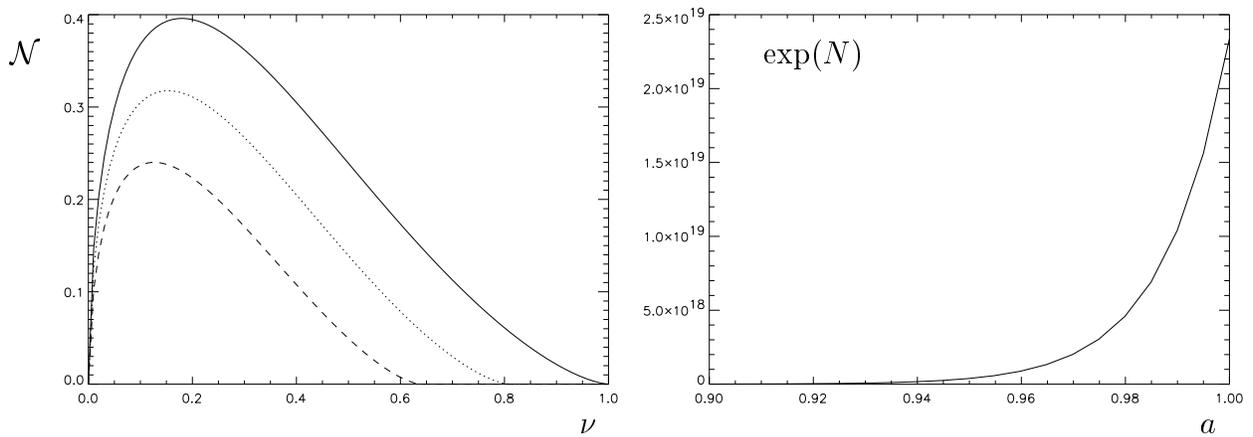}
\caption{\small {\bf Left}:  The growth function $\mathcal{ N}(a,\nu )$ as
a function of $\nu$ at the values of the scale factor $a=0.8$, $a=0.9$ and
$a=1.0$ (dashed, dotted and solid lines, respectively). Clearly,
$\mathcal{ N}$ has a maximum at $\nu \simeq 0.2$ and grows considerably as
the scale factor approaches its present value $a=1$. {\bf Right}: The
growth factor $\exp[N(a,\nu )]$ as a function of the scale factor $a$ at
$\nu = \nu_{\mathrm{max}} \simeq 0.2$. The function $\exp[N(a,\nu)]$
becomes comparable to its present value only at $a \gtrsim 0.95$.
Calculations are done for $w_{p} = -1$, $\sqrt{\alpha }M/H_{0}=100$.
\label{fig1}
}
\end{figure}
as shown in the left panel of Fig.~\ref{fig1}. This is especially relevant
since the gravitational potential (\ref{gravpot}) depends on $N$
exponentially. We present in the right panel of Fig.~\ref{fig1} the
dependence of $\exp[N(a, \nu)]$ on the scale factor at $\nu =0.2$; the
plot is given for the values of parameters\footnote{The effect of the
tachyonic instability on the CMB anisotropy is sizeable if $\sqrt{\alpha
}M/H_0 \sim 100$ (see section \ref{sub:li-exp}), hence our choice here.}
such that $\sqrt{\alpha }M/H_0 =100$. It is clear that the major effect of
tachyonic modes occurs at late times.

\begin{figure}[htb]
\begin{center}
\includegraphics[]{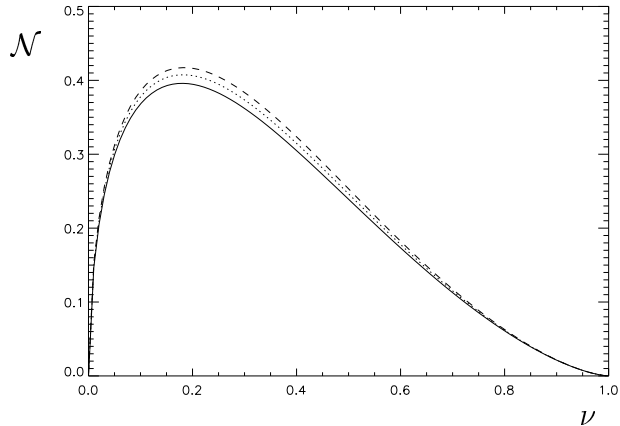}
\end{center}
\caption{\small Function  $\mathcal{ N}
(a,\nu) $ at $a=1$ for different values of the equation of state parameter
of phantom energy, $w_p = -1.0$, $w_p= -1.17$ and $w_p=-1.33$ (solid,
dotted and dashed lines, respectively). }
\label{fig3}
\end{figure}

Finally, let us discuss the dependence of $\mathcal{ N}(a, \nu)$ on the
parameter $w_{p}$. We show in Fig.~\ref{fig3} the function $\mathcal{
N}(a,\nu)$ at $a=1$ for different values of this parameter. Clearly, the
dependence on $w_{p}$ is weak; in what follows we  take $w_{p}=-1$ for
definiteness.

The growth function $\mathcal{ N}(a, \nu)$ as a function of $\nu $ has a
pronounced maximum at $\nu _{\mathrm{max}}=0.2$. The existence of a
maximum of $\mathcal{ N}(a, \nu)$ is a fairly generic property independent
of the particular form of the parametrization (\ref{dispers}), provided
that $\omega ({\bf p})$ vanishes at ${\bf p}=0$. Indeed, the tachyonic
regime begins at $k/a(t)=M$. This implies, in particular, that $\mathcal{
N}(a, \nu)$ vanishes at $\nu =1$. Now, under the assumption that  $\omega
({\bf p}=0) = 0$ one observes that $\mathcal{ N}(a, \nu)$ vanishes at $\nu
=0$ as well. Hence, $\mathcal{ N}(a, \nu)$ has a maximum at some
intermediate $\nu$. In other words, modes of high conformal momentum have
not entered the tachyonic regime yet. On the other hand, $\mathcal{ N}(a,
\nu)$ is small also at low conformal momenta, since the tachyonic regime
begins too early for the corresponding modes. At that time the scale
factor increases too rapidly; the function $\mathcal{ N}(a, \nu)$  does
not have enough time to grow by the epoch when the physical momentum
$k/a(t)$ gets close to zero and the growth terminates. The maximum growth
occurs for intermediate momenta. We note in passing that the situation is
different in the Lorentz-invariant model (\ref{Eq/Pg3/1:PhanTach}), since
in that case $\omega ({\bf p}=0) = M \neq 0$, and the maximum growth takes
place at the lowest spatial momenta.

Besides the growth function $N$, the gravitational potential
(\ref{gravpot}) is determined by the amplitude $A(\mathbf{k})$ at the time
the perturbations  enter the tachyonic regime. So, we have to estimate
this amplitude at $k/a = M$. We assume that the tachyonic perturbations
are generated from vacuum fluctuations. Hence, the amplitude  is the
Gaussian random field with zero expectation value, $\langle A({\bf k})
\rangle = 0$. This field is completely determined by its two-point
correlation function, which we parametrize as follows:
\begin{equation}
\langle A({\bf k}) A^*({\bf k}^{\prime} ) \rangle = \dfrac{f(k)}{k^3}
\frac{H_{0}^2 \delta X^{2}}{M_{PL}^{2}M^{2}} \delta \left( {\bf k} - {\bf
k^{\prime}} \right),
\label{Eq/Pg7/1:PhantomP}
\end{equation}
where $\delta X$ is the primordial perturbations  of the phantom field.
Let us clarify the form chosen for the overall factor. We recall that the
gravitational potential is generated by the perturbations of the phantom
field, which in turn is assumed to cause the accelerated cosmological
expansion. This implies, in particular, that the energy density of the
homogeneous phantom field is of the order of the present critical density
\begin{equation}
\rho _{p}\sim M_{PL}^{2}H_{0}^{2}.
\label{Eq/Pg8/1:Phantom}
\end{equation}
On the other hand, this energy density can be estimated as
\begin{equation}
\rho _{p}\sim M_{X}^{2}X^{2},
\label{Eq/Pg8/2:Phantom}
\end{equation}
where $M_{X}$ is the mass scale characteristic of the phantom field and
$X$ estimates the value of this field. By comparing
eq.~(\ref{Eq/Pg8/1:Phantom}) to eq.~(\ref{Eq/Pg8/2:Phantom}) we obtain an
estimate for this value,
\begin{equation}
X\sim \frac{M_{PL}H_{0}}{M_{X}}.
\label{Eq/Pg13/1:tachyon}
\end{equation}
Let us assume that all dimensionful parameters in the phantom Lagrangian
are of order of the mass parameter $M$ entering the dispersion relation
(\ref{dispers}), so that $M_{X}\sim M$. Therefore,
\[
\delta \rho _{p}\sim M_{X} ^{2}X\delta X\sim M M_{PL}H_{0}\delta X.
\]
Now, we make use of the Poisson equation for the gravitational
potential,
\[
\triangle\Phi \sim \frac{\delta \rho _{p}}{M_{PL}^{2}},
\]
and find that at physical momenta of order $M$ the gravitational
potential is of order
\[
\Phi \sim \frac{H_{0}\cdot\delta X}{M_{PL}\cdot M} \; .
\]
This gives rise to the overall factor in (\ref{Eq/Pg7/1:PhantomP}). The
dimensionless function $f(k)$ in (\ref{Eq/Pg7/1:PhantomP}) parametrizes
possible deviations from the flat spectrum; it is likely that this
function is model dependent. However, we have seen that the growth factor
$\exp[N(a, \nu)]$ is peaked at $\nu = \nu_{max}$. This implies that the
 integrals over conformal momenta, determining CMB anisotropies, are
saturated in a narrow region of $k$. Hence, if $f(k)$ is sufficiently
smooth, our final results are insensitive to its shape, so we can set
$f(k)=C$ where $C$ is a constant. This constant may be somewhat different
from 1, as it may contain, e.g., a power of $H_{0}/M$; it cannot, however,
contain extra factors involving the Planck mass. We will see that our
final results are determined by the interplay between the large growth
factor $\mbox{exp}(N)$ and small factor $\propto M_{PL}^{-1}$ in $A({\bf
k})$, so a possible deviation of $C$ from unity is unimportant in the end.
Furthermore, we will see that the shape of the tachyonic contribution to
the CMB anisotropy spectrum (the position of the maximum and width) is
almost independent of $M$; the parameter $M$, as well as $\delta X$,
determine the overall magnitude only. Therefore, the constant $C$ can be
set equal to 1 by redefinition of the parameter $M$. In view of these
observations we write for the two-point correlator, without loss of
generality,
\begin{equation}
\langle A({\bf k}) A^*({\bf k}^{\prime} ) \rangle = \frac{H_{0}^2 \delta
X^{2} }{M_{PL}^{2}M^{2}} \dfrac{1}{k^3} \delta \left( {\bf k} - {\bf
k^{\prime}} \right) \; .
\label{Eq/Pg7/1:Phantom}
\end{equation}
It is this expression that will be used in what follows. Note that the
amplitude $A(\mathbf{k})$ is  small, so the linearized treatment of the
problem is legitimate even for exponentially growing gravitational
potential.

Let us now discuss possible range of the amplitudes $\delta X$. The lowest
value is obtained by assuming that the perturbations of the phantom field
are in the vacuum state just before the tachyonic regime sets in. Since we
are interested in momenta of order $M$ at that time, their amplitude is
estimated as $\delta X\sim M$. The largest conceivable primordial
perturbations are those generated at inflationary epoch: their amplitude
is then comparable to the amplitudes of other (nonconformal) light fields,
$\delta X\sim H_{\mathit{infl}}/(2\pi )$, where $H_{\mathit{infl}}$ is the
Hubble parameter some 60 $e$-foldings before the end of
inflation~\cite{book1, book2, book3}. In view of the observational
constraint $H_{\mathit{infl}}<1\cdot 10^{-5}M_{PL}$, we estimate the
maximum value of $\delta X$ as $\delta X\sim 10^{-5}M_{PL}$. We will see
that the interesting range of the parameter $M_{X}\sim M$ is roughly
$M\lesssim 10^{3}H_{0}$, so the latter amplitude is small compared to the
background value (\ref{Eq/Pg13/1:tachyon}) of the phantom field.

In what follows we present the results in the two extreme cases,
\begin{equation}
\frac{H_{0}\delta X}{MM_{PL}}\sim \frac{M}{M_{PL}}
\label{a}
\end{equation}
and
\begin{equation}
\frac{H_{0}\delta X}{MM_{PL}}\sim 10^{-5}\frac{H_{0}}{M}.
\label{b}
\end{equation}
The first case is obtained for vacuum amplitudes of fluctuations $\delta
X$ before they get tachyonic, and we neglected the relatively mild
difference between $M$ and $H_{0}$. The second case corresponds to the
generation of primordial phantom perturbations at inflation with the
maximum possible $H_{\mathit{infl}}$. We will see that the fact that these
values differ by many orders of magnitude, the properties of the
contributions to the CMB spectrum are qualitatively similar in the two
cases.

\subsection{Lorentz-invariant model.}
\label{sub:li-growth}

The analysis performed in section~\ref{sub:lv-growth} is straightforwardly
repeated in the case of Lorentz-invariant dispersion relation
(\ref{Eq/Pg3/1:PhanTach}). There are two properties, however, that make
this case different from the Lorentz-violating one. The first property has
to do with the behaviour of the function $\mathcal{ N}(a,\nu )$ that
enters the growth factor,
\begin{equation}
N(a,\nu )=\frac{M}{H_{0}}\mathcal{ N}(a,\nu ) \; .
\label{Eqn/Pg10/1A:PhanTach}
\end{equation}
In the Lorentz-invariant case one has
\begin{equation}
\mathcal{ N}(a,\nu )=\int\limits_{\nu }^a d\hat a\dfrac{\sqrt{\hat
a^{2}-\nu^{2}}}{\sqrt{\hat a}\sqrt{\Omega _{m} + \Omega _{p} \hat
a^{-3w_p}}} \; .
\label{Eqn/Pg10/1:PhanTach}
\end{equation}
Unlike in the Lorentz-violating model, $\mathcal{ N}(a, \nu )$ as a
function of $\nu$ monotonously decreases as $\nu$ increases, so the
maximum of  $\mathcal{ N}(a, \nu )$ is at $\nu=0$. It is therefore
instructive to find $\mathcal{ N}(a, \nu )$ at small $\nu$. To this end,
we write for the integral (\ref{Eqn/Pg10/1:PhanTach})
\begin{equation}
\mathcal{ N}=\mathcal{ N}_{0}+\mathcal{ N}_{1}+\mathcal{ N}_{2},\nonumber
\end{equation}
where
\begin{eqnarray}
\mathcal{ N}_{0}&=&\int\limits_{\nu }^a d\hat
a\dfrac{\hat{a}}{\sqrt{\hat{a}}\sqrt{\Omega _{m} + \Omega _{p} \hat
a^{-3w_p}}},
\label{Eqn/Pg10/2:PhanTach}\\
\mathcal{ N}_{1}&=&\int\limits_{\nu }^\infty d\hat
a\dfrac{\sqrt{\hat{a}^{2}-\nu ^{2}}-\hat{a}}{\sqrt{\hat{a}}\sqrt{\Omega
_{m} + \Omega _{p} \hat a^{-3w_p}}},
\label{Eqn/Pg10/3:PhanTach}\\
\mathcal{ N}_{2}&=&-\int\limits_{a }^\infty d\hat
a\dfrac{\sqrt{\hat{a}^{2}-\nu ^{2}}-\hat{a}}{\sqrt{\hat{a}}\sqrt{\Omega
_{m} + \Omega _{p} \hat a^{-3w_p}}}.
\label{Eqn/Pg10/4:PhanTach}
\end{eqnarray}
The integrals (\ref{Eqn/Pg10/3:PhanTach}), (\ref{Eqn/Pg10/4:PhanTach}) are
convergent, since at large $\hat{a}$ one has
\[
\sqrt{\hat{a}^{2}-\nu ^{2}}-\hat{a}\simeq-\frac{\nu ^{2}}{2\hat{a}}.
\]
Furthermore, the latter expression shows that at small $\nu$ the
integral (\ref{Eqn/Pg10/4:PhanTach}) is of order $\nu ^{2}$. The integral
(\ref{Eqn/Pg10/3:PhanTach}) can be written as
\[
\mathcal{ N}_{1}=\nu ^{3/2}\int \limits_{1}^{\infty }d y
\frac{\sqrt{y^{2}-1}-y}{\sqrt{y}\sqrt{\Omega _{m}+\Omega _{p}\nu
^{-3w_{p}}y^{-3w_{p}}}}.
\]
Since $w_{p}<0$, the term with $\Omega _{p}$ in the integrand can be
neglected at small $\nu $, and then the remaining integral is
straightforwardly evaluated. Finally, the integral $\mathcal{ N}_{0}$ is
readily calculated at\footnote{Like in the Lorentz-violating case, the
dependence on $w_p$ is weak.} $w_{p}=-1$,
\begin{eqnarray}
\mathcal{ N}_{0}&=&\int \limits_{\nu
}^{a}d\hat{a}\frac{\sqrt{\hat{a}}}{\sqrt{\Omega _{m}+\Omega
_{p}\hat{a}^{3}}}=\frac{2}{3\sqrt{\Omega
_{p}}}\left(\mbox{Arcsinh}\sqrt{a^{3}\frac{\Omega _{p}}{\Omega _{m}}}  -
\mbox{Arcsinh}\sqrt{\nu ^{3}\frac{\Omega _{p}}{\Omega _{m}}} \right)
\nonumber\\
&=&\frac{2}{3\sqrt{\Omega _{p}}}\,\mbox{Arcsinh}\sqrt{a^{3}\frac{\Omega
_{p}}{\Omega _{m}}}  - \frac{2}{3\sqrt{\Omega _{m}}}\nu ^{3/2} +\mathcal{
O}(\nu ^{9/2})\; . \nonumber
\end{eqnarray}
With all contributions included, we obtain finally that at small $\nu$
\begin{eqnarray}
\mathcal{ N}(a,\nu )=\frac{2}{3\sqrt{\Omega
_{p}}}\,\mbox{Arcsinh}\sqrt{a^{3}\frac{\Omega _{p}}{\Omega _{m}}}
-\frac{\mathcal{ C}}{\sqrt{\Omega _{m}}}\nu ^{3/2}+\mathcal{ O}(\nu ^{2}),
\label{Eqn/Pg11/1:PhanTach}\\
\mathcal{ C}=\frac{2}{3}-\int \limits_{1}^{\infty
}dy\frac{\sqrt{y^{2}-1}-y}{\sqrt{y}} =\sqrt{\pi }\frac{\Gamma
(-3/4)}{\Gamma (-1/4)}\simeq 1.75.
\end{eqnarray}
Note that the expressions (\ref{Eqn/Pg10/1A:PhanTach}) and
(\ref{Eqn/Pg11/1:PhanTach}) imply that the tachyonic contribution is
non-negligible only for $M>H_{0}$. This is the analog of the inequality
(\ref{Eq/Pg6/1:Phantom}).

It is clear from eq.~(\ref{Eqn/Pg11/1:PhanTach}) that unlike in the
Lorentz-violating case, the growth factor $\mbox{exp}[N(a,\nu)]$ is peaked
at $\nu = 0$. This is due to the fact that low-momentum modes are
tachyonic already at early times, and their ``frequency'' $\omega=M$ is
not small. However, the modes of very low spatial momenta $k =M\nu$
contribute to the {\it  monopole} CMB harmonic only. This contribution
strongly depends on the primordial spectrum of the tachyonic
perturbations, i.e., on the shape of the function $f(k)$ in
(\ref{Eq/Pg7/1:PhantomP}); this is the second special property of the
Lorentz-invariant model. Now,  the monopole contribution merely
renormalizes the average CMB temperature; it is not directly measurable
and will not be discussed in this paper. Multipoles with $l\neq 0$ are
less model-dependent: we will see in section~\ref{sub:li-multi} that the
integrals over momenta are saturated in a relatively narrow region where
$N(a,\nu)$ is not damped, i.e., in the region where
\begin{equation}
\nu \sim \left(\frac{H_{0}}{M}   \right)^{2/3} \; .
\label{Eq/Pg11/1:PhanTach}
\end{equation}
We assume that $f(k)$ does not change much in an interval $\Delta k \sim
k$ for momenta belonging to the region (\ref{Eq/Pg11/1:PhanTach}). Then
for calculating the CMB multipoles with $l\neq 0$, we can still use the
spectrum  (\ref{Eq/Pg7/1:Phantom}). The above discussion of the primordial
amplitude $\delta X$ holds in the Lorentz-invariant case as well. So, we
will again concentrate on the two extreme cases (\ref{a}) and (\ref{b}).

\section{CMB multipoles}
\label{sec:multipoles}

The contribution to CMB anisotropy we are interested in is generated
fairly recently, when the tachyon-induced gravitational potential becomes
sizeable. Therefore, the only phenomenon responsible for this contribution
is the integrated Sachs-Wolfe effect. It is clear from Fig.~\ref{fig1}
that in the Lorentz-violating model this effect operates at the late
cosmological epoch beginning at $z\sim 0.05$ ($a \sim 0.95$). A similar
picture holds in the Lorentz-invariant case.

Let us use the standard notation for the temperature anisotropy
\[
\Theta (\mathbf{n})=\frac{T(\mathbf{n})-T_{0}}{T_0} \; ,
\]
where ${\bf n}$ is the direction of observation and $T_0$ is the
average CMB temperature at present. The integrated Sachs-Wolfe effect then
reads (see, e.g., Ref.~\cite{Giov})
\begin{equation}
\Theta (\mathbf{n})=2\int \limits_{0}^{\eta _{0}}\!d\eta
\left.\frac{\partial \Phi (\eta ,\mathbf{x})}{\partial \eta
}\right|_{\mathbf{x}=\mathbf{n}(\eta _{0}-\eta )},
\label{Eq/Pg9/2:Phantom}
\end{equation}
where  $\eta_0$ is the present time, and in view of the above discussion
we set the lower limit of integration equal to zero instead of the time of
last scattering. The integrand in (\ref{Eq/Pg9/2:Phantom}) exponentially
grows toward the present epoch.

CMB anisotropy is characterized by the multipoles
\begin{equation}
C_l = \dfrac{1}{2l+1} \sum\limits_{m = -l}^{m = l} \langle
|a_{lm}|^2\rangle ,
\label{Eq/Pg9/3:Phantom}
\end{equation}
where $a_{lm}$ are the coefficients of the decomposition of the anisotropy
over spherical harmonics,
\begin{equation}
a_{lm} = \int \; d{\bf n} \; \Theta ({\bf n}) \; Y_{lm}({\bf n}).
\label{multip}
\end{equation}
Making use of eqs.~(\ref{gravpot}), (\ref{Eq/Pg7/1:Phantom}) and
(\ref{Eq/Pg9/2:Phantom}) and performing the angular integration one
obtains the following expression for the multipoles in terms of integrals
over conformal momenta:
\begin{equation}
C_l = \dfrac{8 H_{0}^2\delta X^{2}}{\pi M^{2}M^2_{PL}} \; \;
\int\limits^1_{0} \; \dfrac{d \nu}{\nu} \Delta^2_l(\nu)
\label{multip2}
\end{equation}
The quantity $\Delta^2_l(\nu)$ is the analog of the power spectrum. It is
expressed  in terms of the integral over conformal time,
\begin{equation}
\Delta_{l}(\nu )=\int \limits_{\eta _{k}}^{\eta _{0}}\!d\tau \omega (\tau
)a(\tau )\exp\left\{N[a(\tau ), \nu] \right\}j_{l}[M\nu (\eta _{0}-\tau
)],
\label{Eq/Pg1/1:appendix}
\end{equation}
where
\[
j_l(x) = \sqrt{\dfrac{\pi}{2 x}}J_{l + \frac{1}{2}}(x)
\]
is the spherical Bessel function of the first kind. The integrals
(\ref{Eq/Pg1/1:appendix}) and (\ref{multip2}) have different properties in
Lorentz-violating and Lorentz-invariant models.

\subsection{Lorentz-violating model.} 
\begin{figure}[htb!]
\begin{center}
\includegraphics[scale=0.95]{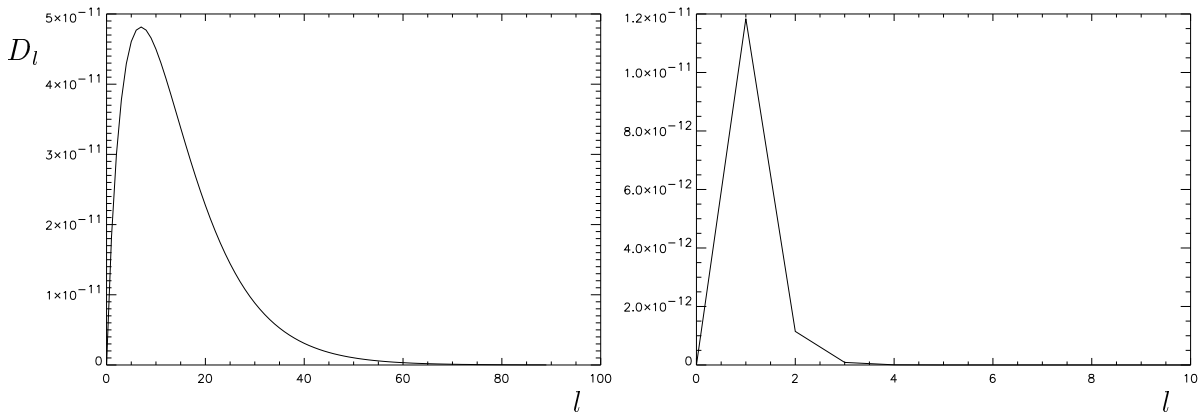}
\end{center}
\caption{\small {\bf Left}:  The tachyonic contribution to the angular
spectrum of CMB temperature in the Lorentz-violating model (\ref{a}) with
$\alpha = 1 \cdot 10^{-3}$ and $M/H_{0} = 9770$. {\bf Right}:  Same, but
for $\alpha = 1.0$ and $M/H_{0} = 315$. }
\label{fig4}
\end{figure}

We plot in the left panel of
Fig.~\ref{fig4} the tachyonic contribution to the angular spectrum of CMB
temperature in the Lorentz-violating model (\ref{a}) with $\alpha = 1
\cdot 10^{-3}$ and $M/H_{0} = 9770$, using the standard quantity
\[
D_l = \dfrac{l(l+1)}{2\pi}C_l \; .
\]
A very similar plot is obtained for the Lorentz-violating model
(\ref{b}) with $\alpha =1\cdot 10^{-3}$, $M/H_{0}=726$. This is a very
general situation: the tachyonic contributions to the CMB spectrum in
models (\ref{a}) and (\ref{b}) at the same value of $\alpha $ are
virtually indistinguishable if the value of $M$ appropriately scaled down
by about an order of magnitude. The reason for that will become clear
later, see eqs. (\ref{Eq/Pg3/1:appendix}), (\ref{Eq/Pg14/1:Phantom}). The
analysis of this section applies to both models (\ref{a}) and (\ref{b});
we will illustrate our results using the model (\ref{a}) for definiteness.

It is clear that the spectrum shown in Fig.~\ref{fig4} has a rather narrow
maximum at $l=l_{\mathrm{max}}$ ($l_{\mathrm{max}} \approx 7$ in this
example). We will see below that the position of the maximum is determined
solely by the parameter $\alpha$, and that $l_{\mathrm{max}}$ grows as
$\alpha$ decreases, see eq.~(\ref{Eq/Pg14/1:Phantom}).

For $\alpha = 1$, the spectrum is shown in the right panel of
Fig.~\ref{fig4}. It is clear from this figure that at large enough
$\alpha$, the tachyonic perturbations contribute to the lowest multipoles
only. It is also clear that the dipole and quadrupole components differ by
one order of magnitude only. Hence, we disregard in what follows the
dipole component, as for realistic quadrupole it is negligible compared to
the observed dipole, which is supposedly due to the motion of the Earth in
the CMB reference frame. We note in passing that the situation is
different in the Lorentz-invariant model: it is meaningful to discuss the
dipole component in that case, see sections \ref{sub:li-multi} and
\ref{sub:li-exp}.

The multipoles $C_l$ can be calculated analytically in the regime $\alpha
(M/H_{0})^{2} \gg 1$,  see (\ref{Eq/Pg6/1:Phantom}). As we discussed in
section~\ref{sub:lv-growth}, the function $\mbox{exp}[N(a,\nu )]$ rapidly
grows with the scale factor, and as a function of $\nu$ it has a peak at
$\nu _{\mathrm{max}}\simeq 0.2$. Thus, the main contribution into the
integral (\ref{Eq/Pg1/1:appendix}) comes from late times, while the
integral (\ref{multip2}) is saturated at $\nu \approx \nu _{\mathrm{max}}$
(it is important at this point that $\nu _{\mathrm{max}}$ is different
from both 0 and 1). This means, in the first place, that the lower limit
of integration in (\ref{Eq/Pg1/1:appendix}) may be set equal to zero.
Second, one can make use of expansion of the function $N(a, \nu)$ near
$a=1$,
\begin{equation}
N[a(\eta ), \nu]=N(1, \nu)-\omega(\nu) \cdot(\eta _{0}-\eta ),
\label{Eq/Pg1/2:appendix}
\end{equation}
and for $\nu=\nu _\mathrm{max}\simeq 0.2$ we have
\[
N(1,\nu _{\mathrm{max}})=0.39 \sqrt{\alpha }\frac{M}{H_{0}}.
\]
Third, one can set the pre-exponential factor $\omega(\eta ) a(\eta )$
in (\ref{Eq/Pg1/1:appendix}) equal to its value at $a=1$. In this way we
obtain
\begin{equation}
\Delta_{l}(\nu )=\sqrt{\alpha }\sqrt{\frac{1-\nu}{\nu } }\exp\left[
N(1,\nu )\right]\int \limits_{0}^{\nu M\eta_{0}
}\!dx\exp\left(-\sqrt{\alpha }\sqrt{\frac{1-\nu }{\nu }}x \right)j_{l}(x),
\label{Eq/Pg1/3:appendix}
\end{equation}
where we introduced the integration variable $x=\nu M(\eta _{0}-\eta )$.
Finally, let us recall that $\eta _{0}\sim 1/H_{0}$. Then the upper limit
of integration in (\ref{Eq/Pg1/3:appendix}) is $\nu M\eta _{0}\sim \nu
_{\mathrm{max}}M/H_{0}\gg 1$, so the integration may be extended to
infinity. Then the integral in (\ref{Eq/Pg1/3:appendix}) is calculated by
making use of the formula~\cite{GR-Abr}
\begin{eqnarray}
\int \limits_{0}^{\infty
}\!dx\frac{J_{l+\frac{1}{2}}(x)}{\sqrt{x}}\exp\left(-\gamma x
\right)&=&\frac{1}{(1+\gamma ^{2})^{\frac{1}{4}}}\Gamma
(l+1)P_{-\frac{1}{2}}^{-l-\frac{1}{2}}\left[\frac{\gamma }{\sqrt{1+\gamma
^{2}}} \right]=\nonumber\\
&=&\frac{1}{(1+\gamma^{2})^{\frac{1}{4}}}\frac{\Gamma (l+1)}{\Gamma
(l+\frac{3}{2})}\left[\frac{1-z}{1+z} \right]^{\frac{l}{2}+\frac{1}{4}}
F\left(\frac{1}{2},\frac{1}{2},l+\frac{3}{2},\frac{1-z}{2}\right),
\label{Eqn/Pg15/1:PhanTach}
\end{eqnarray}
where $P_{-\frac{1}{2}}^{-l-\frac{1}{2}}$ is the Legendre function,  $F$
is the hypergeometric function\footnote{The second representation of the
integral (\ref{Eqn/Pg15/1:PhanTach}) is convenient because the
hypergeometric function is only slightly different from unity at $0 \leq z
\leq 1$, which corresponds to $0\leq \alpha <\infty $.}, and
\[
z=\frac{\gamma}{ \sqrt{1+\gamma ^{2}}} =\sqrt{\frac{\alpha (1-\nu
)}{\nu +\alpha (1-\nu )}}.
\]
Because of the exponential dependence on $\nu $ of the factor $\exp
[N(1,\nu )]$ in (\ref{Eq/Pg1/3:appendix}), the integral in (\ref{multip2})
can be evaluated in the saddle-point approximation, and we obtain
\begin{eqnarray}
D_l &=&\frac{2H_{0}^{2}\delta X^{2}
}{M^{2}M_{PL}^{2}\nu_{\mathrm{max}}}\sqrt{\frac{1}{\pi |N''(1,\nu
_{\mathrm{max}})|}}\exp(2N(1,\nu _{\mathrm{max}})) \cdot l(l+1)\cdot\left[
\frac{\Gamma (l+1)}{\Gamma (l+3/2)}\right]^{2}\times\nonumber\\
&&\times \frac{z^{2}}{\sqrt{1-z^{2}}}\left(\frac{1-z}{1+z}
\right)^{l+1/2}F^2
\left(\frac{1}{2},\frac{1}{2},l+\frac{3}{2},\frac{1-z}{2} \right) \; ,
\label{exact}
\end{eqnarray}
where $z=z(\nu _{\mathrm{max}})$, $N''(1,\nu )$ is the second derivative
with respect to $\nu $. At $\nu _{\mathrm{max}} \simeq 0.2$ we have
\begin{eqnarray}
D_l & = & \mathcal{A}_0 \cdot
\left[\dfrac{\Gamma(l+1)}{\Gamma(l+3/2)}\right]^2 \dfrac{l(l+1)}{\sqrt{1 +
4\alpha}} \cdot \left(\dfrac{1-z}{1+z}\right)^{l+1/2}
F^2\left(\frac{1}{2}, \frac{1}{2}; l+\frac{3}{2}, \frac{1-z}{2}\right),
\label{exact1}\\
z&=&\dfrac{2\sqrt{\alpha}}{\sqrt{1 + 4\alpha}},\nonumber\\
\mathcal{A}_0 &=& 9.4 \cdot \alpha ^{3/4}\left(\frac{H_{0}}{M}
\right)^{5/2}\frac{\delta X^{2}}{M_{PL}^{2}} \cdot
\exp\left[0.78\sqrt{\alpha}\frac{M}{H_0}\right].
\label{C0value}
\end{eqnarray}
This is the desired analytical expression for the multipoles; it works
with the precision which is certainly sufficient for our purposes..

At $\alpha \ll 1$ and $l>1$, the expression (\ref{exact}) simplifies to
\begin{equation}
D_l=\frac{2\alpha H_{0}^{2}\delta X^{2} }{M^{2}M_{PL}^{2}}\frac{1-\nu
_{\mathrm{max}}}{\nu _{\mathrm{max}}^{2}}\sqrt{\frac{1}{\pi |N''(1,\nu
_{\mathrm{max}})|}}\exp(2N(1,\nu _{\mathrm{max}})) \cdot
(l+1)\cdot\exp\left(\!-2l\sqrt{\alpha }\sqrt{\frac{1-\nu_{\mathrm{max}}
}{\nu_{\mathrm{max}} }} \right).
\label{Eq/Pg3/1:appendix}
\end{equation}
Corrections to the latter formula are   $\mathcal{ O}(1/l)$ and they are
numerically small even for $l=2$.

A few comments are in order. First, it is seen from (\ref{exact1}) that
the dependence on $l$ and on $M$ has factorized. Therefore, the position
of the maximum $l_{\mathrm{max}}$ and the width of the peak in the
spectrum depend on the parameter $\alpha $ and do not depend on the
parameters $M$ and $\delta X$. In particular, for $\alpha \ll 1$ the
maximum of the function (\ref{Eq/Pg3/1:appendix}) is at
\begin{equation}
l_{\mathrm{max}}=\frac{1}{2\sqrt{\alpha }}\sqrt{\frac{\nu
_{\mathrm{max}}}{1-\nu _{\mathrm{max}}}}-1\simeq \frac{1}{4\sqrt{\alpha
}}-1.
\label{Eq/Pg14/1:Phantom}
\end{equation}
The overall magnitude of the spectrum depends on  $\alpha $, $M$ and
$\delta X$. The dependence on $M$ is exponential, since $N(1,\nu
_{\mathrm{max}}) \propto M$. This justifies the use of
(\ref{Eq/Pg7/1:Phantom}) for the primordial spectrum of perturbations of
the gravitational potential.

Second, the exponential dependence of $C_{l}$ on $l$ has the following
interpretation. The problem has the characteristic time scale $\tau
(k)\sim \omega^{-1} =(\sqrt{\alpha }M\sqrt{\nu (1-\nu )})^{-1} \ll
H^{-1}_{0}$. This scale determines the time of the development of the
tachyonic instability in a mode with momentum $k$. Since the growth
function $N(\nu )$ has a maximum, the relevant modes have momenta near
$k_{\mathrm{max}}=\nu _{\mathrm{max}}M$. Therefore, the gravitational
potential is small at  distances $r\gg \tau (k_{\mathrm{max}})$ along the
light cone emanating from the observer (the tachyonic instability has not
developed yet). On the other hand, at distances $r<\tau
(k_{\mathrm{max}})$ the gravitational potential is almost constant in time
(according to (\ref{Eq/Pg6/1:Phantom}), the expansion of the Universe has
a negligible effect, while the tachyonic instability gives rise to mild
growth of the potential). In other words, the gravitational potential at
$r< \tau (k_{\mathrm{max}})$ is the superposition of random,
time-independent waves with almost constant amplitude and almost constant
wavelength $2\pi/k_{\mathrm{max}}$. At $r> \tau (k_{\mathrm{max}})$ the
amplitude of these waves decays as $\exp[-r/\tau(k _{\mathrm{max}})]$ as
$r$ increases. The period of a wave located at distance $r$ is seen at an
angle $\triangle\theta _{r}\simeq 2\pi/(rk_{\mathrm{max}})$. Hence, this
wave contributes to the multipoles with $l\simeq rk_{\mathrm{max}}$. The
multipoles $a_{lm}$ are not exponentially suppressed for $r<\tau
(k_{\mathrm{max}})$, i.e., $l < 1/[k_{\mathrm{max}}\tau
(k_{\mathrm{max}})]$, and are exponentially small in the opposite case.
Recalling (\ref{Eq/Pg9/3:Phantom}), one finds that this behaviour of
$a_{lm}$ leads to the following dependence of $C_l$ on $l$,
\[
C_{l}\propto \exp\left(-\frac{2l}{k_{\mathrm{max}}\tau
(k_{\mathrm{max}})} \right)= \exp\left(-2l\sqrt{\alpha }\sqrt{\frac{1-\nu
_{\mathrm{max}}}{\nu _{\mathrm{max}}}} \right) \; ,
\]
in complete agreement with (\ref{Eq/Pg14/1:Phantom}).

The fact that for relatively large $\alpha $ sizeable contributions are
obtained by the lowest multipoles only (see Fig.~\ref{fig4}) can be seen
directly from (\ref{Eq/Pg9/2:Phantom}). Indeed,  inserting (\ref{gravpot})
into (\ref{Eq/Pg9/2:Phantom}), making use of (\ref{Eq/Pg1/2:appendix}),
and integrating over time, we obtain
\begin{equation}
\Theta (\mathbf{n})\sim\int \limits_{}^{}\!d^{3}k F(k,\omega )
A(\mathbf{k}) \frac{1}{\sqrt{\alpha }\sqrt{\frac{1-\nu }{\nu }}
-i\frac{\mathbf{k \cdot n}}{k}}+\mbox{h.c.},
\label{nov29-1}
\end{equation}
where $F(k,\omega )$ is a smooth function independent of the direction of
$\mathbf{k}$. Recalling that $\nu \simeq \nu _{\mathrm{max}}=0.2$, we have
$\sqrt{\alpha }\sqrt{(1-\nu)/\nu } >1$ for sufficiently large $\alpha $.
In that case the denominator in (\ref{nov29-1}) can be expanded in a
series in ($\mathbf{k \cdot n}$), which just corresponds to the expansion
in spherical harmonics. The $l$-th harmonic is thus suppressed as $
(\sqrt{\alpha }\sqrt{(1-\nu)/\nu })^{-l} $. This is relatively mild
suppression, in accord with the right panel of Fig.~\ref{fig4}. At
$\sqrt{\alpha }\sqrt{(1-\nu)/\nu } <1$ the expansion of the denominator in
(\ref{nov29-1}) is not legitimate, and one has to perform more
 sophisticated analysis leading to (\ref{exact1}).

\subsection{Lorentz-invariant model.}
 \label{sub:li-multi}

In the Lorentz-invariant case we make use of the expression
(\ref{Eqn/Pg11/1:PhanTach}) for the function $\mathcal{ N}(a,\nu )$ to
calculate the integral (\ref{Eq/Pg1/1:appendix}). We find
\begin{equation}
\Delta_{l}(\nu )=\frac{\sqrt{\pi }}{2}\frac{\Gamma (l+1)}{\Gamma
(l+3/2)}\left(\frac{\nu }{2}
\right)^{l}\cdot\exp\left(\frac{2}{3}\frac{M}{H_{0}\sqrt{\Omega
_{p}}}\,\mbox{Arcsinh}\sqrt{\frac{\Omega _{p}}{\Omega
_{m}}}-\frac{\mathcal{ C}M}{H_{0}\sqrt{\Omega _{m}}}\nu ^{3/2} \right).
\label{Eq/Pg17/1:PhanTach}
\end{equation}
When obtaining this expression we  used the fact that $\nu$ is small (see
(\ref{Eq/Pg11/1:PhanTach})), again extended the time integration to
infinity in the same way as we have done after
eq.~(\ref{Eq/Pg1/3:appendix}) and made use of in
eq.~(\ref{Eqn/Pg15/1:PhanTach}). We also kept the leading terms in $\nu$
in the expression (\ref{Eq/Pg17/1:PhanTach}).

Inserting (\ref{Eq/Pg17/1:PhanTach}) into the integral in the expression
(\ref{multip2}) for multipoles, and changing the integration variable, we
arrive at the following integral:
\begin{equation}
\left(\frac{H_{0}\sqrt{\Omega _{m}}}{2\mathcal{ C}M} \right)^{4l/3}\int
\limits_{0}^{\frac{\mathcal{ C}M}{H_{0}\sqrt{\Omega _{m}}}}dx
x^{4l/3-1}\mathrm{e}^{-x}=\left(\frac{H_{0}\sqrt{\Omega _{m}}}{\mathcal{
C}M} \right)^{4l/3} \gamma \left(\frac{4l}{3}, \frac{\mathcal{
C}M}{H_{0}\sqrt{\Omega _{m}}}  \right),
\label{Eq/Pg18/1:PhanTach}
\end{equation}
where $\gamma(\beta  ,x) $ is an incomplete $\Gamma $ function. We notice
that the second argument of this function
\[
x = \frac{\mathcal{ C}M}{H_{0}\sqrt{\Omega _{m}}} \; ,
\]
is large, $x \gg 1$, otherwise the overall factor in (\ref{multip2})
makes the effect we discuss negligibly small. Hence, for $\beta \equiv
4l/3 \ll x$ we use the approximation $\gamma(\beta  ,x) = \Gamma (\beta)$,
   and obtain the final formula for the multipoles
\begin{equation}
D_l = \frac{H_{0}^2\delta X^{2}}{3\pi
M^{2}M^2_{PL}}\frac{l(l+1)}{2^{2l-1}} \exp
\left(\frac{4M}{3H_0}\frac{1}{\sqrt{\Omega_p}}\mbox{Arcsinh}
\sqrt{\frac{\Omega_p}{\Omega_m}}\right) \left[
\dfrac{\Gamma(l+1)}{\Gamma(l+3 / 2)}\right]^2\! \left( \dfrac{H_0
\sqrt{\Omega_m}}{2\mathcal{ C}M}\right)^{4l \over 3}
\!\!\!\!\Gamma\!\!\left (\frac{4l}{3}\right).
\label{power1}
\end{equation}
This formula is valid for relatively low multipoles, $l<\mathcal{
C}M/(H_{0}\sqrt{\Omega _{m}})$. In the opposite case the behaviour of the
incomplete $\Gamma$ function is
\[
\gamma (\beta ,x)\simeq \frac{x^{\beta }}{\beta }\mathrm{e}^{-x}\,
\]
Hence, the multipoles at large $l$ are negligibly small in the
interesting case of large $x$.

Because of the exponential factor in (\ref{power1}), the low multipoles
may be fairly large at large enough $M/H_0$. On the other hand, a large
value of this parameter implies that the multipoles rapidly decay with
$l$. Consider first the case (\ref{a}) of small primordial amplitude
$\delta X$. As we discuss in section \ref{sub:li-exp}, the interesting
range is $M/H_0 \sim 100$, in which case the multipole $D_{l+1}$ is
suppressed by a factor of about $10^{-3}$ as compared to $D_l$. So, it is
sufficient to consider the dipole and quadrupole anisotropies only. With
our parametrization, these are, respectively,
\begin{eqnarray}
D_1 &=& 6.4\cdot 10^{-125} \left(\dfrac{M}{H_0}\right)^{2/3} \exp
\left(1.98\cdot \frac{M}{H_0}\right)
\label{Eqn/Pg19/1:PhanTach}\\
D_2 &=& 4.1\cdot 10^{-126} \left(\dfrac{H_0}{M}\right)^{2/3} \exp
\left(1.98\cdot \frac{M}{H_0}\right)
\label{Eqn/Pg19/2:PhanTach}\\
\dfrac{D_1}{D_2} &=& 15.7 \left(\dfrac{M}{H_0}\right)^{4/3}
\label{Eqn/Pg19/3:PhanTach}
\end{eqnarray}
These expressions are in agreement, within $10\%$, with the values that we
obtained numerically\footnote{The reason for the $10\%$ discrepancy is the
omission of the term of order  $\nu^{2}$ in the expression
(\ref{Eqn/Pg11/1:PhanTach}) for $\mathcal{N}(a,\nu)$; note that
$\mathcal{N}(a,\nu)$ enters  the final result exponentially.}.

Let us now turn to high primordial amplitude $\delta X$, case (\ref{b}).
In that case, the interesting range is $M/H_{0}\sim 10$, and
\[
D_2 = 2.7\cdot 10^{-14}  \left(\dfrac{H_0}{M}\right)^{14/3} \exp
\left(1.98\cdot \frac{M}{H_0}\right),
\]
while eq. (\ref{Eqn/Pg19/3:PhanTach}) remains valid.

\section{Comparison with the data}
\label{sec:data}

Overall, the data on the anisotropy of CMB temperature are in good
agreement with the standard picture of adiabatic scalar perturbations
whose primordial spectrum is close to the Harrison-Zeldovich one. Still,
the observed angular spectrum may possibly show deviations whose nature is
unclear. Our analysis was partially motivated by the desire to understand
whether these deviations may be due to the contributions coming from the
tachyonic perturbations of phantom energy. As we discuss in this section,
the deviations {\it cannot} be explained in this way. So, our analysis
enables us only to place limits on the parameters of the tachyonic
perturbations.

\subsection{Lorentz-violating case.}

We begin with the Lorentz-violating model, and consider a wide range of
the parameter $\alpha$,
\begin{equation}
2.5\cdot 10^{-7} < \alpha < 1.0.
\label{Eq/Pg14/1:Phantom15_09}
\end{equation}
This range is representative: at $\alpha \gtrsim 1$ the main effect is in
the lowest multipoles, whereas at $\alpha =2.5\cdot 10^{-7}$ the
contribution is peaked at $l\sim l_{\mathrm{max}}\simeq 500$, see
(\ref{Eq/Pg14/1:Phantom}). As we pointed out above, we do not consider the
dipole anisotropy in the Lorentz-violating case, as it is much smaller
than the observed dipole anisotropy supposedly originating from  the
motion of the Earth.

In our study we used the data on multipoles given in Ref.~\cite{wmapcl}
and organized as a table ``$C_l$ vs. $l$.'' These are not combined in
bins, unlike the data usually presented.

In the model we study, the multipoles are the sums of two terms, one due
to adiabatic scalar perturbations and another due to tachyonic modes,
\begin{equation}
C_l = C_l^{(ad)} + C_l^{(t)} \; .
\label{dec2-1}
\end{equation}
Making use of the code CMBFast~\cite{cmbfast}, we calculated the angular
spectrum $C_l^{(ad)}$ generated by adiabatic scalar perturbations for
various values of the spectral index $n_s$ in the range $0.8 \le n_s \le
1.5$, in the standard cosmological model with the following values of
 parameters: the Hubble constant $H_0 = 72~\mbox{km} \cdot \mbox{s}^{-1}
\cdot \mbox{Mpc}^{-1}$, baryon plus CDM contribution to the present energy
density $\Omega_{m} = 0.27$, contribution of hot dark matter $\Omega_{hdm}
= 0$, dark energy contribution $\Omega_p \equiv \Omega_{\Lambda} = 0.73$,
$^4\mbox{He}$ abundance $Y=0.24$, number of massless neutrino species
$N_\nu=3$. We assumed that the tensor perturbations are absent. The second
term $C_l^{(t)}$ in (\ref{dec2-1}) was calculated by making use of the
analytical expression (\ref{exact1}).

To compare the model with the data, we analyzed the difference between the
measured and calculated multipoles,
\[
\epsilon_l = C_l^{(exp)} - C_l^{(ad)} - C_l^{(t)}.
\]
The study of the moments and correlation properties of $\epsilon_l$ has
shown that they are independent and their average is zero within
statistical error. Furthermore, an $\chi^2$ estimate has shown that   with
$95\%$ probability the distribution of $\epsilon_l$ is Gaussian.

To obtain the parameters of the theoretical spectrum, we used maximum
likelihood method with the likelihood function
\begin{eqnarray}
F( \epsilon | \theta) &=& \prod\limits_{l=2, 600} f(\epsilon_l |\theta) \;
, \nonumber \\
f(\epsilon_l |\theta) &=& \exp\left(-\dfrac{\epsilon_l^2(\theta)}{2
\sigma_l^2}\right) \; . \nonumber
\end{eqnarray}
Here $\theta$ is the set of four parameters: the spectral index and
amplitude of  adiabatic perturbations,  the amplitude $\mathcal{A}_0$ and
 parameter $\alpha$ of the tachyonic perturbations. In view of
(\ref{Eq/Pg14/1:Phantom}) and (\ref{Eq/Pg14/1:Phantom15_09}), we included
in our analysis multipoles with $l\leq 600$ only. For each $\alpha$ from
the range (\ref{Eq/Pg14/1:Phantom15_09}) we obtained the best fit values
of the three other parameters. The variations of individual multipoles
have been calculated with the use of the error estimations given in the
third column of the table in Ref.~\cite{wmapcl}.

\begin{figure}[htb!]
\begin{center}
\includegraphics[]{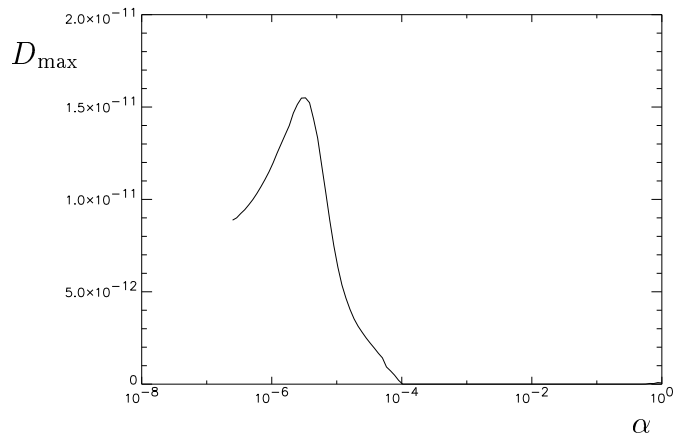}
\end{center}
\caption{
\small The best fit value of the tachyonic contribution to CDM temperature
anisotropy as a function of $\alpha$. The parameter $D_{\mathrm{max}}$ is
defined in (\ref{nov30-1}). The maximum of $D_{\mathrm{max}}$ is at
$\alpha= 3.0 \cdot 10^{-6}$, which corresponds to $l_{\mathrm{max}}\simeq
143$.
\label{fig6}
}
\end{figure}

We show in Fig.~\ref{fig6} the best fit value of the tachyonic
contribution as a function of the parameter $\alpha$. As a measure of this
contribution we use the maximum in the anisotropy spectrum generated by
the tachyonic perturbations,
\begin{equation}
D_{\mathrm{max}} = \max_{l} \left[ \frac{l(l+1)}{2\pi} C^{(t)}_l \right]
\; .
\label{nov30-1}
\end{equation}
This maximum is at $l=l_{\mathrm{max}}$ (as an example, $l_{\mathrm{max}}
\approx 7$ in the left panel of Fig.~\ref{fig4}). It is clear from
Fig.~\ref{fig6} that at $\alpha > 10^{-4}$ the best fit value is equal to
zero, whereas at $\alpha < 10^{-4}$ it is considerably different from
zero. This means that the tachyonic contribution improves the agreement
between the theory and data. It is worth noting that the addition of this
contribution moves the best fit value of the spectral index up from
$n_{s}=0.96$ obtained in Ref.~\cite{wmapcos}; in particular, for some
values of $\alpha$ the best fit values of $n_s$ are larger than 1.

\begin{figure}[tb]
\begin{center}
\includegraphics[]{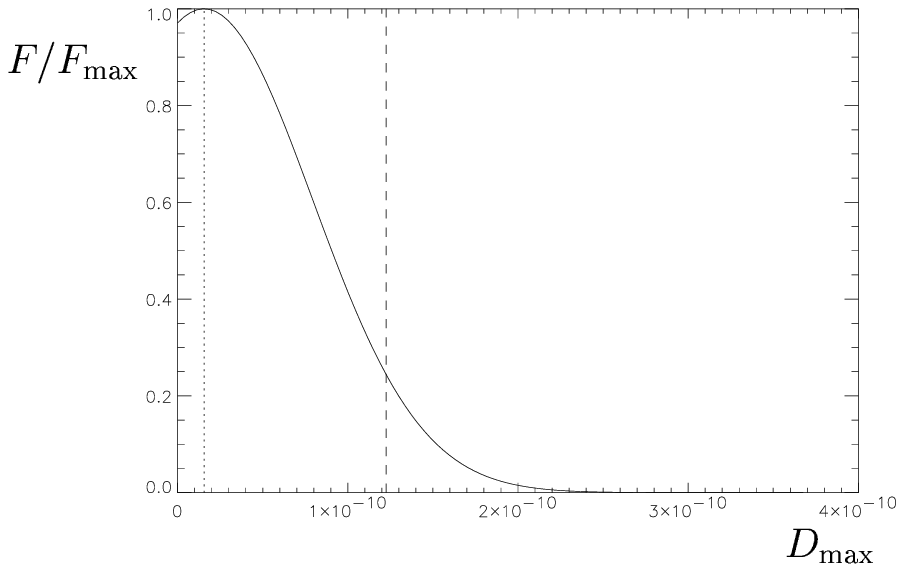}
\end{center}
\caption{\small
Likelihood function at the best fit values of the amplitude
and spectral index of the adiabatic perturbations as
a function of $D_{\mathrm{max}}$ at
$\alpha = 1.8 \cdot 10^{-6}$. The dashed line
shows the limit on  $D_{\mathrm{max}}$ at
95~\% confidence level.
The dotted line corresponds to the best fit value of
$D_{\mathrm{max}}$.
The maximum value of the likelihood function is
$F_{\mathrm{max}} = 0.577$.
}
\label{fig7}
\end{figure}

This improvement, however, is not statistically significant. We show in
Fig.~\ref{fig7} the maximum likelihood function as a function of
$D_{\mathrm{max}}$ at $\alpha=1.8 \cdot 10^{-6}$. It is clear from
Fig.~\ref{fig7}, that even though the best fit value of $D_{\mathrm{max}}$
is nonzero, the difference of the likelihood function at the best fit
value and at $D_{\mathrm{max}}=0$ is small. The same is true for all
values of $\alpha$ in the range considered. So, the data is consistent
with the absence of the tachyonic contribution to the CMB temperature
anisotropy.

\begin{figure}[tb]
\begin{center}
\includegraphics[]{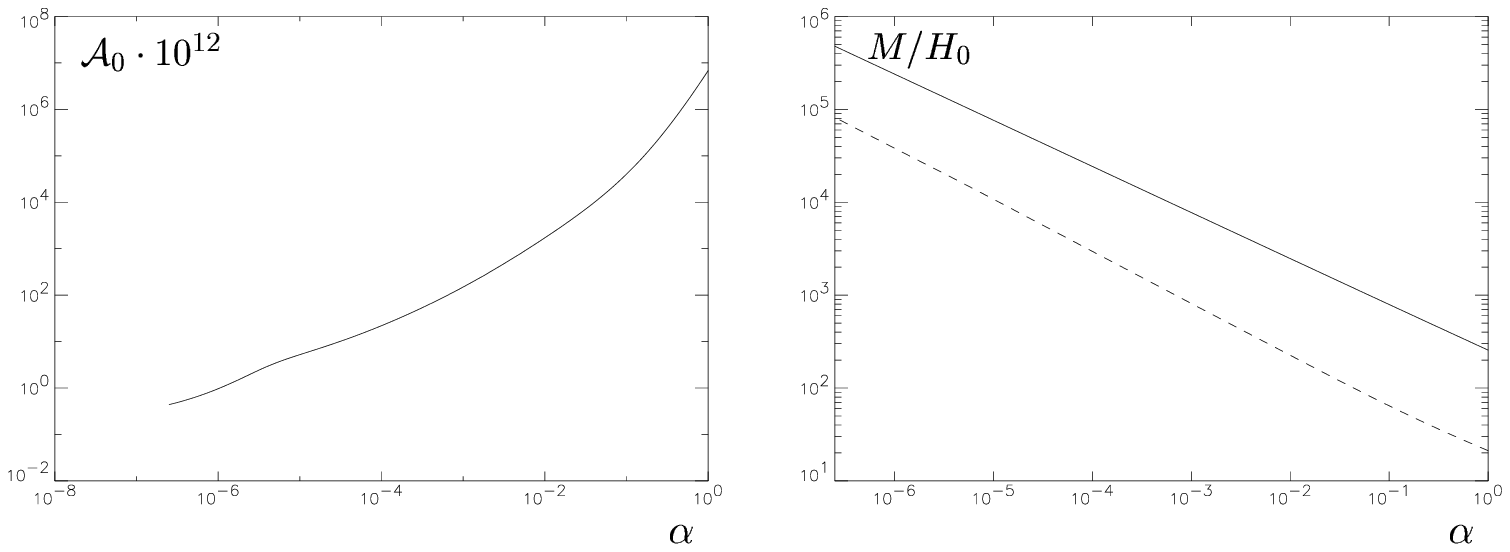}
\end{center}
\caption{\small {\bf Left}:
Upper limit on the amplitude $\mathcal{A}_0$ for $\alpha \in\left[1.0
\div 2.5\cdot 10^{-7}\right]$ at 95~\% confidence level. {\bf Right}:
Upper limit on the parameter $M$ of the tachyonic perturbations in units
of the Hubble constant, at 95~\% confidence level, in the model (\ref{a})
(solid line) and model (\ref{b}) (dashed line). }
\label{fig8}
\end{figure}

Thus, we can only place limits on the overall magnitude  of the tachyonic
contribution, $\mathcal{A}_0$, at various values of $\alpha$, which can
then be translated into the limits on the physical parameter $M/H_0$.
These limits, at $95~\% $ confidence level, are shown in Fig.~\ref{fig8}
for both cases of low primordial amplitude (\ref{a}) and high amplitude
(\ref{b}).

\subsection{Lorentz-invariant case.}
\label{sub:li-exp}

As we have seen in section~\ref{sub:li-multi}, in the model with
Lorentz-invariant spectrum the CMB multipoles generated by the tachyonic
perturbations rapidly decrease as $l$ increases. Therefore, only two
multipoles --- the dipole and quadrupole --- are relevant for comparison
with the data. The measured dipole component of CMB temperature
is~\cite{wmap03_1} $d = 3.358 \pm 0.017$ mK, and the direction in the
Galactic polar coordinate frame is  $l = 263.86 \pm 0.04^{\circ}$,
$b=48.24 \pm 0.10^{\circ}$. In the standard parametrization, the dipole
anisotropy is
\begin{equation}
D^{exp}_1 = \frac{1}{3\pi}\sum\limits_{m = -1}^{m = 1} |a_{1 m}|^2=
1.6\cdot 10^{-7},
\label{dip}
\end{equation}
while the quadrupole component is given by
\begin{equation}
D^{exp}_2 = \frac{3}{5\pi}\sum\limits_{m = -2}^{m = 2} |a_{2 m}|^2 =
2.9\cdot 10^{-11}.
\label{quad}
\end{equation}
To obtain conservative limits on the parameter $M$ of Lorentz-invariant
tachyonic perturbations, we do not impose any priors on the contribution
to the dipole anisotropy due to the motion of the Earth and on the
quadrupole anisotropy generated by adiabatic perturbations. In the model
with small primordial spectrum (\ref{a}) we make use of the expression
(\ref{Eqn/Pg19/1:PhanTach}) and take into account the fact that the dipole
has 3  degrees of freedom. In this way we find that the analysis of the
dipole anisotropy  leads to the limit
\begin{equation}
\frac{M}{H_{0}}\le 135.9 \;\;\;\;\; \mbox{at}~~95~\%~~\mbox{c.l.}\;.
\label{Eq/Pg24/1:PhanTach}
\end{equation}
The limit coming from the quadrupole anisotropy is obtained by making use
of eq.~(\ref{Eqn/Pg19/2:PhanTach}). It reads
\begin{equation}
\frac{M}{H_{0}}\le 136.4\;\;\;\;\; \mbox{at}~~95~\%~~\mbox{c.l.}\;.
\label{Eq/Pg24/2:PhanTach}
\end{equation}
Interestingly,  the limit coming from the dipole anisotropy is similar to
that obtained from the quadrupole. One can turn this result around and
speculate that the large observed dipole  may be due to the tachyonic
perturbations, with no contradiction to the data at higher multipoles. In
our model the latter property is natural in the sense that the quadrupole
and higher angular harmonics are small automatically. Another way to
phrase this is to pretend  that the observed dipole anisotropy is due to
the tachyonic perturbations, i.e., equate (\ref{Eqn/Pg19/1:PhanTach}) and
(\ref{dip}), and then calculate the contribution to the quadrupole from
 (\ref{Eqn/Pg19/2:PhanTach}). This gives for the tachyonic contribution
$D_2 = 1.5\cdot 10^{-11}$, which is safely below  the observed value
(\ref{quad}). The octupole is suppressed by another three orders of
magnitude. So, our model would serve as an alternative to other
explanations~\cite{GZ,Turner,PP,JP,lan96, lan96a,lan97} of the large
dipole component of the CMB anisotropy, if such an explanation were needed
(for observational aspects of this issue see
Refs.~\cite{Kamionkowski:2002nd,Gordon:2007sk} and references therein).

In the model with large primordial perturbations (\ref{b}), the situation
is different. In that case, the strongest limit on $M/H_{0}$ is obtained
from the quadrupole
\[
\frac{M}{H_{0}}<8.6\;\;\;\;\; \mbox{at}~~95~\%~~\mbox{c.l.}\;.
\]
As seen from eq. (\ref{Eqn/Pg19/3:PhanTach}), the dipole is not so much
enhanced as compared to quadrupole; at $M/H_{0}=8.6$ its value is
$D_{1}\simeq 1\cdot 10^{-8}$, which is safely below the observed value
(\ref{dip}). On the other hand, the octupole and higher harmonics are
still suppressed compared to the quadrupole by more than an order of
magnitude.

\section{Discussion}
\label{sec:conclusion}

In this paper we have considered the effects on the anisotropy of CMB
temperature due to possible tachyonic perturbations of dark energy.
Because of the exponential growth, these perturbations may generate large
gravitational potential $\Phi$ at the recent cosmological epoch, and only
at that epoch. This results in a sizeable Sachs-Wolfe effect. Note that
the tachyonic perturbations we have discussed are unrelated to
perturbations in baryons or dark matter, so their contribution to the CMB
anisotropy does not correlate with the distribution of structure in the
Universe.

Our analysis was mostly motivated by the Lorentz-violating models of
phantom energy. Hence, we have studied in detail the tachyonic
perturbations with the dispersion relation (\ref{dispers}). We have seen
 that their effect on the CMB angular spectrum has a pronounced maximum
whose position depends on one of the parameters, $\alpha$, and is
practically insensitive to  other parameters. It is expected that similar
shape of the angular spectrum is characteristic to a wide class of models
with Lorentz-violating tachyonic perturbations, as it is closely related
to the fact that these perturbations become sizeable at late times only.

We have also considered tachyonic perturbations with Lorentz-invariant
dispersion relation (\ref{Eq/Pg3/1:PhanTach}). In that case, an
interesting possibility is that the largest contribution to the CMB
ansotropy is received by the dipole component, and the angular spectrum
rapidly decays with the increase of $l$. We have seen that even if the
entire observed dipole anisotropy is attributed to the tachyonic
perturbations, the quadrupole component generated by them is still
consistent with the observational data. It is worth noting that this
result should be inherent not only in the Lorentz-invariant model, but
also in Lorentz-violating models with tachyonic dispersion relations,
 provided that the ``frequency'' does not vanish at zero momentum and
decreases as momentum increases.

Our main conclusion is that even if perturbations of the tachyonic type
exist in the Universe, their contribution to the CMB anisotropy is small.
Nevertheless, we do not exclude a possibility that growing precision of
observations, and especially elaborate analysis of correlations between
CMB anisotropy and structures in the Universe, may lead to hints toward
the possible exotic property of dark energy, the tachyonic behaviour of
its perturbations.

\section*{Acknowledgemets} This work has been partially supported by
Russian Foundation for Basic Research grant Nos. 07-02-01034a (O.S. and
M.S) and 08-02-00473 (M.L. and V.R), grant of the President of RF for
leading scientific schools under Contract No. NS-1616.2008.2 (M.L and
V.R.), grant of the President of RF under Contract No. MK-2503.2008.2
(O.S.) and grant of Dynasty Foundation (M.L.) M.L. is indebted to
Universit\'e Libre de Bruxelles, where part of this work has been done
under partial support  by the Belspo:IAP-VI/11 and IISN grants, for
hospitality.

\end{document}